\documentstyle[11pt]{article}

\begin{document}
\author{ Jen-Fa Min  \thanks{min@ibm65.phys.ncku.edu.tw}
 \\
Department of Physics, National Cheng Kung 
University, Tainan,Taiwan, 70101, R.O.C
 \and Huei-Shih Liao  \thanks{d847301@phys.nthu.edu.tw}
 \\
 Department of Physics, National Tsing Hua 
University, Hsin-Chu, Taiwan, 30043, R.O.C. }
\title{Studies on the Chiral Order Parameter in the Schwinger Model
}

\maketitle

\begin{abstract}
Based on an analytical technique using a unitary transformation and the
variational method, we study the chiral order parameter in the Schwinger 
model in the lattice formalism with Kogut-Susskind fermions. The fermion
condensate $\langle \stackrel {-}{\Psi }\Psi \rangle $ 
for any coupling constant and fermion mass 
are calculated. Chiral symmetry is shown to be broken in the massless 
limit and good scaling behavior is obtained.

\newpage\ 
\end{abstract}

\input tcilatex

\section{Introduction}

\hspace{6mm}In the field of lattice gauge theory (LGT), there are two major
directions of studies. One is to create new techniques such that Monte-Carlo
(MC) simulation for systems with fermions so as to get exact
results efficiently. The other is to investigate the universality of lattice
action and the existence of the continuum limit of LGT.

The Schwinger model\cite{1}, which describes 1+1 dimensional $QED,$ is
superrenomalizable and exactly solvable\cite{1} -- \cite{6}.
Previous investigations of the model reveal some important
properties of $QCD$ such as quark confinement, chiral symmetry breaking and $%
U(1)$ problem. LGT with fermions in 1+1 dimensions is easier to study
than in higher dimensions. The study of lattice formalism of the Schwinger
model is a good test of the two major directions of LGT. Before we apply
the technique to more realistic systems, the studies of the Schwinger model 
can give us a better understanding of LGT.

The investigations of the Schwinger model have been carried out by 
MC simulation and by other analytic approach\cite{7} -- \cite{14}. 
Although MC simulation gives
several exciting results, analytic methods are still necessary to realize
the quatitative picture of its lattice formalism. However, even the
finite-lattice technique\cite{15} -- \cite{18}, one of the
most successful analytic methods, is restricted to the strong coupling
regime. New analytic approaches still need to be explored.

Recently, a technique with naive fermion formalism\cite{19} is developed. It
contains a unitary transformation and the variational method such that the
Hamiltonian can be diagonalized and reliable physical informations can be
extracted from the Hamiltonian LGT. Our study is concentrated on the
Kogut-Susskind formalism of this approach for the Schwinger model. The
chiral order parameter $\langle \stackrel{-}{\Psi }\Psi \rangle $ is
calculated and its scaling behavior is investigated.

This paper is organized as follows. In Section $2$, we give a brief review
of the Kogut-Susskind formalism for the Schwinger model. The unitary
transformation and the variational method are introduced in Section $3$.
Finally,
the results and discussion are given in Section $4$.

\section{The Schwinger model on a lattice }

\hspace{6mm}We consider the Schwinger model on a lattice and use the 
Hamiltonian formalism with Kogut-Susskind fermions. In this formalism,
the upper and 
lower components of the fermion spinor are put on different lattice sites. 
We shall work in the temporal gauge and the Hamiltonian reads 
 
\begin{equation}
\label{1}H=-\frac 1{2a}\stackunder{x}{\sum }\left[ \chi ^{+}(x)U(x,x+1)\chi
(x+1)+h.c.\right] +m_0\stackunder{x}{\sum }(-1)^x\chi ^{+}(x)\chi (x)+\frac{%
g^2}{2a}\stackunder{x}{\sum }L^2(x),
\end{equation}
where $\chi ^{+}(x)$ $(\chi (x))$ is a fermionic raising (lowering) operator at
site $x$, $U (x,x+1)$ is the gauge field on the link between site $x$ and $x+1$,
$a$ is the lattice spacing, $m_0$ is the bare mass of the fermion and $g$ is the
dimensionless bare coupling constant related to the charge by $g=ae$. $U(x,x+1)$
can be written as 

\begin{equation}
\label{2}U(x,x+1)=e^{iaeA(x)}=e^{i\theta (x)}=b^{+}(x).
\end{equation}
The dimensionless operator $L(x)$ which generates rotation in $\theta (x)$ is 
the canonical conjugate to $\theta (x)$: 

\begin{equation}
\label{3}\left[ L(x),e^{\pm i\theta (y)}\right] =\pm \delta _{x,y}e^{\pm
i\theta (x)}.
\end{equation}
Specifying $\left| n\right\rangle$ as the eigenstate of $L (x)$, we see that $b^+
(x)$ $(b (x))$ raises (lowers) the boson number by one:
 
\begin{equation}
\label{4}
\begin{array}{l}
b^{+}\left| n\right\rangle =\left| n+1\right\rangle,  \\ 
b\left| n\right\rangle =\left| n-1\right\rangle.
\end{array}
\end{equation}
We also need the commutation relations between the boson operators and the fermion
operators:

\begin{equation}
\label{5}\left[ \chi ^{+}(x),b^{+}(x)\right] =\left[ \chi
(x),b^{+}(x)\right] =\left[ \chi ^{+}(x),b(x)\right] =\left[ \chi
(x),b(x)\right] =0.
\end{equation}
The fermion operators $\chi$ and $\chi^{+}$ satisfy the usual algebra:

\begin{equation}
\label{6}\left\{ \chi ^{+}(x),\chi (y)\right\} =\delta _{x,y},\left\{ \chi
^{+}(x),\chi ^{+}(y)\right\} =\left\{ \chi (x),\chi (y)\right\} =0.
\end{equation}
\hspace{6mm}If we consider the dimensionless parameter $g^{2}$ to be large (strong 
coupling), the first term in the Hamiltonian can be considered as a perturbed term.
The second and the last terms are the unperturbed terms. The ground state $\left| 
0\right\rangle$ of the unperturbed Hamiltonian has the following properties:

\begin{equation}
\label{7}
\begin{array}{l}
L^2(x)\left| 0\right\rangle =0,
\text{ for all }x, \\ \chi ^{+}(x)\chi (x)\left| 0\right\rangle =\{
\begin{array}{c}
0,
\text{ for even }x \\ \left| 0\right\rangle ,\text{ for odd }x.
\end{array}
\end{array}
\end{equation} 
Because odd lattice sites are occupied by an antiparticle, this state 
corresponds to a $``$filled Dirac sea."

The gauge invariant states are specified by fermion occupation numbers 
and by boson numbers. They are denoted by 

\begin{equation}
\label{8}\left| \Psi \right\rangle =\stackunder{x}{\prod }(\chi
^{+}(x))^{f(x)}(b^{+}(x))^{s(x)}\left| 0\right\rangle
, 
\end{equation}
where $\left| 0\right\rangle$ is the $``$filled Dirac sea," and $f(x)$, $s(x)$
are the
numbers of fermions and bosons at location $x$, respectively.

\section{Unitary transformation and the variational method}

\hspace{6mm}We use the unitary transformation:

\begin{equation}
\label{9}H^{^{\prime }}=\exp (-i\theta S_f)H\exp (i\theta S_f)
,
\end{equation}
to diagonalize the Hamiltonian. 
The physical vacuum is assumed to be

\begin{equation}
\label{10}\left| \Omega \right\rangle =\exp (i\theta S_f)\left|
0\right\rangle
,
\end{equation}
where

$$
S_f=\frac i{\sqrt{2}}\stackunder{x}{\sum }(-1)^x\left[ \chi
^{+}(x)U(x,x+1)\chi (x+1)-\chi ^{+}(x+1)U^{+}(x,x+1)\chi (x)\right]  
,
$$
and $\theta$ is the variational parameter.

The physical vacuum energy is then given by

\begin{equation}
\label{11}E_\Omega =\left\langle 0\right| H^{^{\prime }}\left|
0\right\rangle =\left\langle 0\right| \exp (-i\theta S_f)H\exp (i\theta
S_f)\left| 0\right\rangle 
.
\end{equation}
Defining

\begin{equation}
\label{12}H^{^{\prime }}=\exp (-i\theta S_f)H\exp (i\theta
S_f)=H_k^{^{\prime }}+H_m^{^{\prime }}+H_g^{^{\prime }}
,
\end{equation}
where

\begin{equation}
\label{13}
\begin{array}{l}
H_k^{^{\prime }}=\exp (-i\theta S_f)\left( -\frac 1{2a}
\stackunder{x}{\sum }\left[ \chi ^{+}(x)U(x,x+1)\chi (x+1)+h.c.\right]
\right) \exp (i\theta S_f) \\ H_m^{^{\prime }}=\exp (-i\theta S_f)\left( m_0
\stackunder{x}{\sum }(-1)^x\chi ^{+}(x)\chi (x)\right) \exp (i\theta S_f) \\ 
H_g^{^{\prime }}=\exp (-i\theta S_f)\left( \frac{g^2}{2a}\stackunder{x}{\sum 
}L^2(x)\right) \exp (i\theta S_f),
\end{array}
\end{equation}
and using the Baker-Hausdorff formula

$$
e^AFe^{-A}=F+\left[ A,F\right] +\frac 1{2!}\left[ A,\left[ A,F\right]
\right] +\frac 1{3!}\left[ A,\left[ A,\left[ A,F\right] \right] \right]
+......... 
,
$$
and the commutation relations, we can work out $H_k^{^{\prime }}$,
$H_m^{^{\prime }}$, and $H_g^{^{\prime }}$.

We first consider $H_m^{^{\prime }}$. In the expansion of
the Baker-Hausdorff formula, we find that, because 
there is one link in $S_f$ and none in $H_m^{^{\prime }}$, there appear odd 
numbers of $U$, $U^{+}$ combinations for the even terms.  
 
\begin{equation}
\left\langle 0\right| U^m(U^{+})^n\left| 0\right\rangle =0\text{ if }m+n
\text{ is odd.} 
\end{equation}

We therefore conclude that

$$\begin{array}{c}
\left\langle 0\right| H_m^{^{\prime }}\left| 0\right\rangle  \\  
\\  
\\ 
\end{array}
\begin{array}{l}
=
\stackunder{n=0}{\sum }\frac 1{(2n)!}(-\sqrt{2}\theta )^{2n}m_0(-1)^n\binom{%
2n}n\stackunder{x}{\sum }(-1)^x\left\langle 0\right| \chi ^{+}(x)\chi
(x)\left| 0\right\rangle  \\ =-m_0
\frac{N_l}2\stackunder{n=0}{\sum }\frac 1{(2n)!}(\sqrt{2}\theta )^{2n}(-1)^n
\binom{2n}n \\ =-m_0\frac{N_l}2J_0(2\sqrt{2\theta }),
\end{array}
$$
where $N_l$ is the total number of the lattice sites, and $J_0 (x)$ is the  
Bessel function of the first kind.

We next consider $H_k^{^{\prime }}$. Because there is one link in $S_f$
and also one in $H_k^{^{\prime }}$, the expansion of the Baker-Hausdorff 
formula shows that for the odd terms there appear odd numbers of $U$, $U^ {+}$ 
combinations. These terms cannot exist in
$\left\langle 0\right| H_k^{^{\prime }}  
\left| 0\right\rangle$. We finally get

\begin{equation}
\label{15}
\begin{array}{c}
\left\langle 0\right| H_k^{^{\prime }}\left| 0\right\rangle  \\  
\\  
\\  
\\ 
\end{array}
\begin{array}{l}
=
\stackunder{n=0}{\sum }\frac 1{(2n+1)!}(-\sqrt{2}\theta )^{2n+1}\frac
1{2a}(-1)^{n+1}\binom{2n+2}{n+1}\stackunder{x}{\sum }(-1)^x\left\langle
0\right| \chi ^{+}(x)\chi (x)\left| 0\right\rangle  \\ =\frac 1{2a}(-
\frac{N_l}2)\stackunder{n=0}{\sum }\frac 1{(2n+1)!}(-\sqrt{2}\theta
)^{2n+1}(-1)^{n+1}\binom{2n+2}{n+1} \\ =\frac 1{2a}(-
\frac{N_l}2)*2\stackunder{n=0}{\sum }\frac 1{n!(n+1)!}(-1)^n(\sqrt{2}\theta
)^{2n+1} \\ =-\frac 1a\frac{N_l}2J_1(2\sqrt{2\theta )},
\end{array}
\end{equation}
with  $J_1(x)$ being the Bessel function of the first kind.

Finally, we consider the last term $H_g^{^{\prime }}$. 
$H_g^{^{\prime }}$ can be rewritten as 

\begin{equation}
\label{16}
\begin{array}{c}
\left\langle 0\right| H_g^{^{\prime }}\left| 0\right\rangle  \\  
\\  
\\ 
\end{array}
\begin{array}{l}
=\left\langle 0\right|\exp (-i\theta S_f)\left( 
\frac{g^2}{2a}\stackunder{x}{\sum }L^2(x)\right) \exp (i\theta S_f) 
\left| 0\right\rangle \\ =
\left\langle 0\right|
\frac{g^2}{2a}\stackunder{x}{\sum }\exp (-i\theta S_f)L(x)\exp (i\theta
S_f)\exp (-i\theta S_f)L(x)\exp (i\theta S_f) \left| 0\right\rangle\\ =
\left\langle 0\right|\frac{g^2}{2a}\stackunder{x%
}{\sum }\left[ L(x)+B(x)+C(x)\right] \left[ L(x)+B(x)+C(x)\right] 
\left| 0\right\rangle ,
\end{array}
\end{equation}
where

\begin{equation}
\label{17}
\begin{array}{c}
\left\langle 0 \right |B(x)\left | 0\right\rangle= \\  
\\ 
\end{array}
\begin{array}{c}
\stackunder{n=0}{\sum }\frac 1{(2n+1)!}(\frac{\sqrt{2}\theta }%
2)^{2n+1}(-1)^{n+1}\stackunder{y}{\sum }(-1)^y\stackunder{m=0}{\stackrel{2n+1%
}{\sum }}\stackunder{m^{^{\prime }}=m-2n}{\stackrel{m}{\sum }}\binom{2n+1}m
\binom{2n}{m-m^{^{\prime }}} \\ \chi ^{+}(y)U^{2n+1-m}(U^{+})^m\chi
(y+2n-2m+1)(-1)^{2n-2m+m^{^{\prime }}+2}\delta _{y,x+m^{^{\prime }}}.
\end{array}
\end{equation}

\begin{equation}
\label{18}
\begin{array}{c}
\left\langle 0 \right |C(x) \left |0\right\rangle= \\  
\\ 
\end{array}
\begin{array}{c}
\stackunder{n=1}{\sum }\frac 1{(2n)!}(\frac{\sqrt{2}\theta }2)^{2n}(-1)^{n+1}%
\stackunder{y}{\sum }\stackunder{m=0}{\stackrel{2n}{\sum }}\stackunder{%
m^{^{\prime }}=m-2n+1}{\stackrel{m}{\sum }}\binom{2n}m\binom{2n-1}{%
m-m^{^{\prime }}} \\ \chi ^{+}(y)U^{2n-m}(U^{+})^m\chi
(y+2n-2m)(-1)^{2n-2m+m^{^{\prime }}+1}\delta _{y,x+m^{^{\prime }}}.
\end{array}
\end{equation}
We now only need to calculate $B^2 (x)$ and $C^2 (x)$ in: 

$$\left\langle 0\right| H_g^{^{\prime }}\left| 0\right\rangle =\frac{g^2}{2a}%
\stackunder{x}{\sum }\left[ \left\langle 0\right| B^2(x)\left|
0\right\rangle +\left\langle 0\right| C^2(x)\left| 0\right\rangle \right]  
.
$$
Here we use the fact that $L(x)\left| 0\right\rangle =0$ and $\left\langle  
0\right|  B(x)C(x)\left|0\right\rangle =0$. The reason for $\left\langle 
0\right|  B(x)C(x)\left|0\right\rangle =0$ is that $\left\langle 0\right|
B(x)C(x)\left|0\right\rangle $ involves
 $\left\langle 0\right| \chi ^{+}(y) 
\chi (y+2n-2m+1)\chi^{+}(y^{^{\prime }})\chi (y^{^{\prime }}+2n^{^{\prime }}
-2m^{^{\prime}})\left| 0\right\rangle $.
The only chance that it will not be zero is that 
$y^{\prime}=y+2n-2m+1$ and $y=y^{\prime}+2n^{\prime}
-2m^{\prime}$. This implies that $2n-2m+1=2m^{\prime}-2n^{\prime}$. This is not
possible because the left hand side is odd and the right hand side is even.
Therefore, we have 

\begin{equation}
\label{19}\frac{g^2}{2a}\stackunder{x}{\sum }\left\langle 0\right|
B^2(x)\left| 0\right\rangle =\frac{g^2}{2a}(\frac{N_l}2)\frac 12\stackunder{%
n=1}{\sum }\frac 1{(2n)!}(\sqrt{2}\theta )^{2n}(-1)^{n+1}\binom{2n}n\binom{%
2n-2}{n-1},
\end{equation}

\begin{equation}
\label{20}
\begin{array}{l}
\frac{g^2}{2a}\stackunder{x}{\sum }\left\langle 0\right| C^2(x)\left|
0\right\rangle  \\  \\  
\\ 
\end{array}
\begin{array}{l}
=
\frac{g^2}{2a}(\frac{N_l}2)[\stackunder{n=2}{\sum }2^{-3}\frac 1{(2n)!}(
\sqrt{2}\theta )^{2n}(-1)^n\binom{2n}n\binom{2n}n \\ \text{ }-\stackunder{n=2%
}{\sum }2^{-2}\frac 1{(2n)!}(\sqrt{2}\theta )^{2n}(-1)^n\binom{2n}n \\ =
\frac{g^2}{2a}(\frac{N_l}2)\stackunder{n=1}{\sum }\frac 1{(2n)!}(\sqrt{2}%
\theta )^{2n}(-1)^n\binom{2n}n\left[ \frac 18\binom{2n}n-\frac 14\right]. 
\end{array}
\end{equation}
We wind up with

\begin{equation}
\label{21}
\begin{array}{l}
\left\langle 0\right| H_g^{^{\prime }}\left| 0\right\rangle  \\  
\\  
\\ 
\end{array}
\begin{array}{l}
=
\frac{g^2}{2a}(\frac{N_l}2)\stackunder{n=1}{\sum }\frac 1{(2n)!}(\sqrt{2}%
\theta )^{2n}(-1)^n\binom{2n}n\left[ \frac 18\binom{2n}n-\frac 14-\frac 12
\binom{2n-2}{n-1}\right]  \\ = 
\frac{g^2}{2a}(\frac{N_l}2)\{{\frac 18J_0^2(2\sqrt{2}\theta )-\frac 14J_0(2
\sqrt{2}\theta )+\frac 18-\frac 14\int_0{}^{2\sqrt{2}\theta
}\int_0^x{}dydx\left[ J_1^2(y)-J_0^2(y) \right] }\} .\\ 
\end{array}
\end{equation}
\hspace{6mm}The dimensionless ground state energy now reads

\begin{equation}
\label{22}
\begin{array}{c}
\epsilon _\Omega =
\frac{aE_\Omega }{N_l} =  \\  \\ 
\end{array}
\begin{array}{l}
-\frac 12m_0aJ_0(2
\sqrt{2\theta })-\frac 12J_1(2\sqrt{2\theta })+\frac{g^2}4\{\frac 18J_0^2(2
\sqrt{2}\theta )-\frac 14J_0(2\sqrt{2}\theta )+\frac 18 \\ 
-\frac 14\int_0^{2
\sqrt{2}\theta }\int_0^xdydx\left[ J_1^2(y)-J_0^2(y) \right].
\end{array}
{}{}
\end{equation}
 
By $\frac{\partial \epsilon _\Omega }{\partial \theta }=0$, we can
determine the parameter $\theta$ for any coupling constant and bare fermion 
mass.

\section{Results and Discussion}

\hspace{6mm}Based on the unitary transformation and the variational method we
addressed in the last section, we have studied the vacuum structure and chiral
symmetry breaking in the Schwinger model with Kogut-Susskind fermions. The
reason we prefer to use the Kogut-Susskind formalism is its 
simplicity.

The fermion condensate $\langle \stackrel{-}{\Psi }\Psi 
\rangle $ can be calculated by

$$\langle \stackrel{-}{\Psi }\Psi \rangle =\left\langle \Omega
\right| \stackunder{x}{\sum }(-1)^x\chi ^{+}(x)\chi (x)\left| \Omega
\right\rangle /N_l=-\frac 12 J_0(2\sqrt{2\theta }).
$$
Figure 1 plots $\langle \stackrel{-}{\Psi }\Psi\rangle
/g $ as a function of $1/ g^2$ in the chiral limit $m_0 = 0$. It shows a 
very good scaling behavior although the result we calculated, $\langle 
\stackrel{-}{\Psi }\Psi\rangle = -0.252 \pm 0.034$, is higher than the
exactly calculated result:

$$\langle \stackrel{-}{\Psi }\Psi \rangle /e=-\exp (\gamma )/2\pi
^{3/2}=-0.16.
$$
Our result would be better if we add the interaction with next to
nearest-neighbor term to $S_f$. Figure 2 shows $\langle \stackrel{-}{\Psi }
\Psi\rangle$ against the bare fermion mass for $1/ g^2 =1.0,2.0,$ and
$3.0$, respectively. 

For quite a long time, LGT cannot escape the restrict of  
strong coupling regime. With the application of a unitary transformation and 
the variational method, the awkward situation of LGT has been overcome. In a 
future work, we will use the technique to study the chiral symmetry properties 
and mass spectrum of a two dimensional $SU(2)$ gauge theory coupled to an 
unflavored Susskind fermion.

This work was supported by the National Science Council of Republic
of China under Grant No. NSC 83-0208-M-006-015. We thank Professor Su-Long 
Nyeo for fruitful discussions.

\newpage\

\newpage\
\\
{\Large \bf Figure Captions}
 
\vspace{1cm}

\noindent
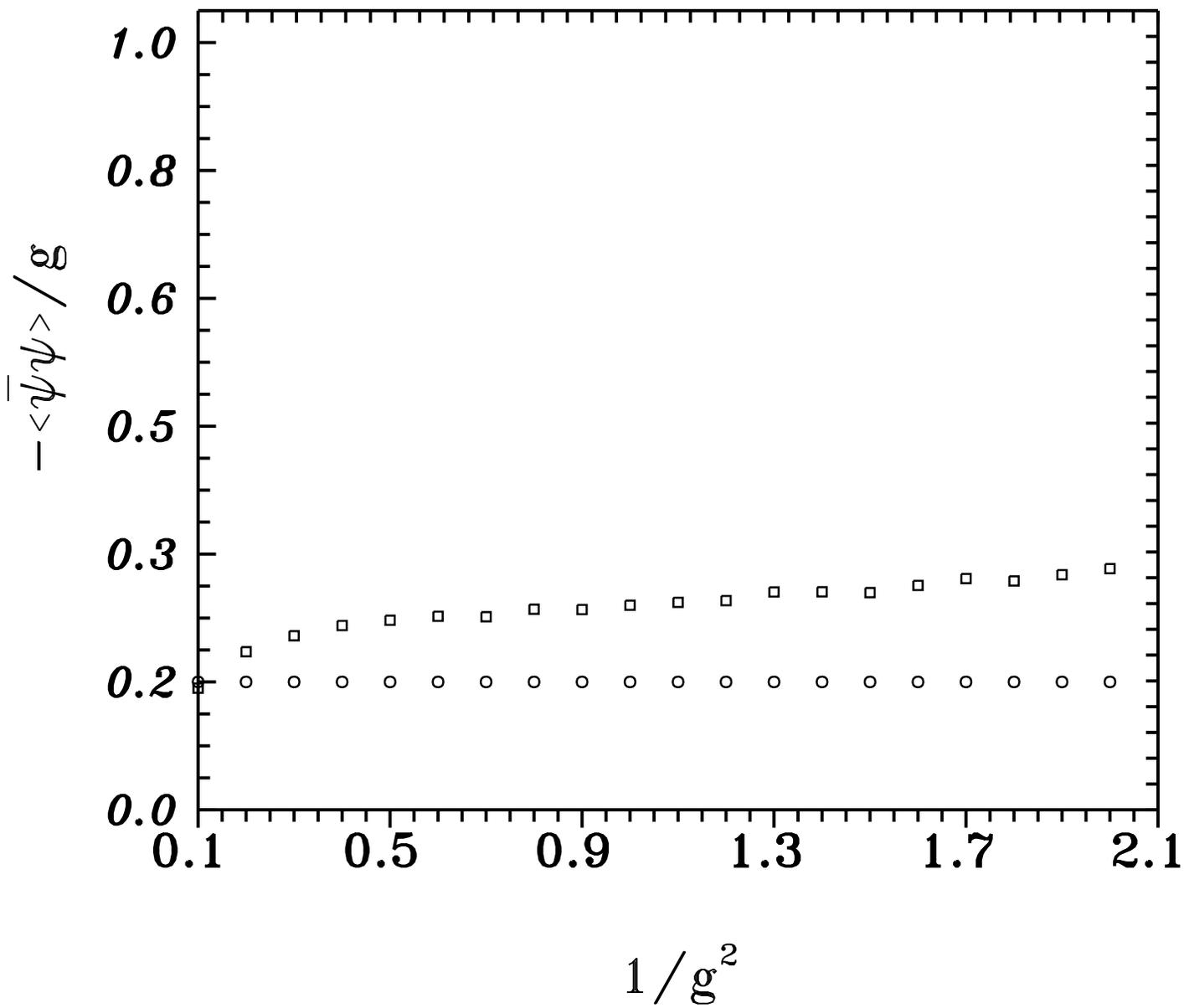
{\bf Fig.1} 
$-\langle \stackrel{-}{\Psi}\Psi \rangle /g$ was plotted as a 
function of $1/g^2$ in the chiral limit $m_0=0$. $g^2$ ranges from $0.5$
to $10$, where the symbols square and circle represent our results and 
the exactly 
calculated value, respectively.
\\
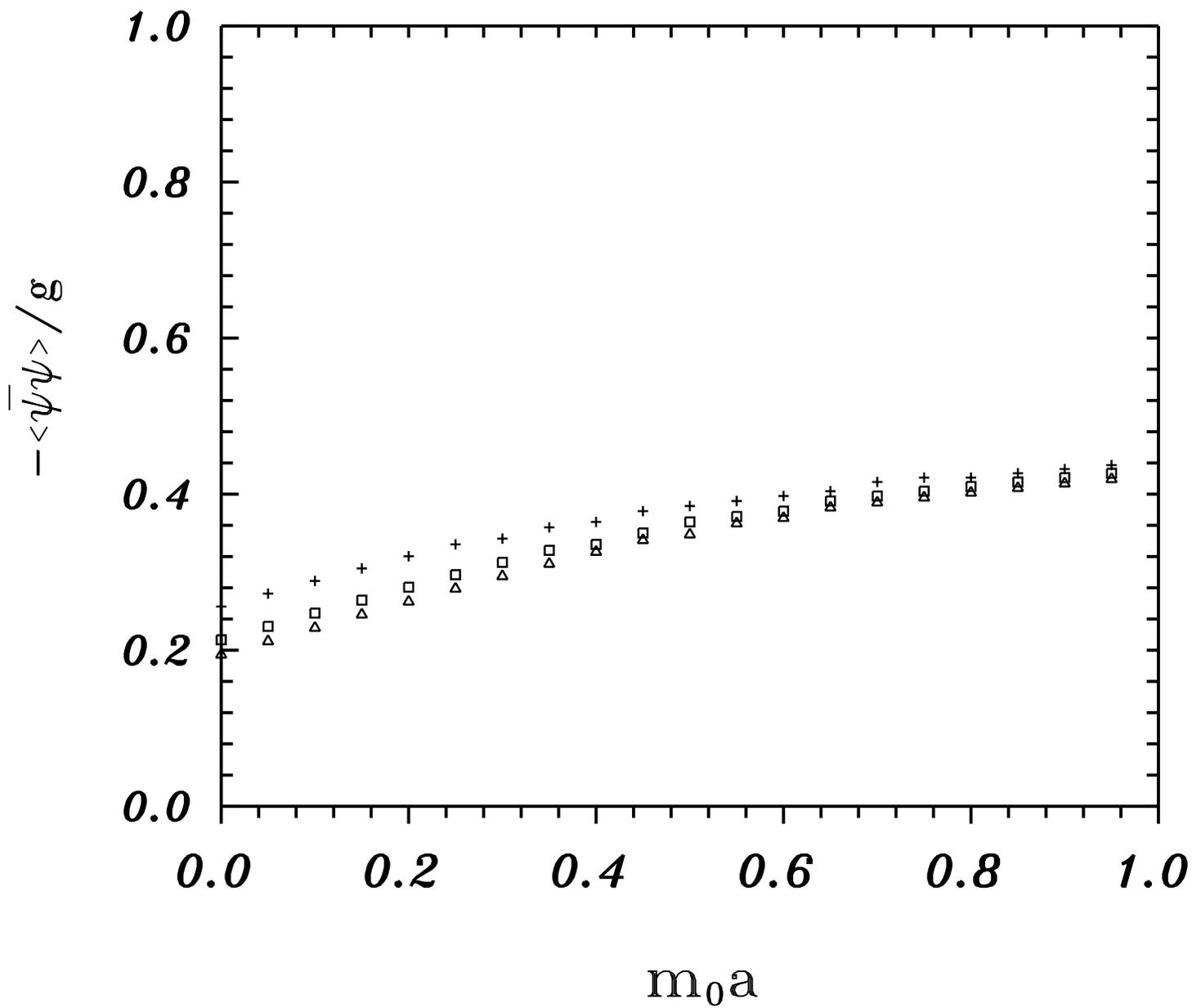
{\bf Fig.2} $-\langle \stackrel{-}{\Psi}\Psi \rangle$ was plottted as a 
function of bare fermion mass $m_0a$ for $1/g^2=1.0,2.0,$ and $3.0$, where
the symbols cross, square, and triangle represent $1/g^2=1.0,2.0,3.0$, 
respectively.
\end{document}